\documentclass{jetpl}
\twocolumn

\usepackage{amsmath,amssymb,epsfig,color,pstricks,bm,graphics,alltt}

\begin{document}

%%% article in English
\lat

%%% article title
\title{Electronic Structure of New LiFeAs High-T$_c$ Superconductor}

%%% article title - for colontitle (at the top of the page)
\rtitle{Electronic Structure of LiFeAs}

%%% article title - for table of contents (usualy identical with \title)
\sodtitle{Electronic Structure of New LiFeAs High-T$_c$ Superconductor}

%%% author(s) ( + e-mail)
\author{I.\ A.\ Nekrasov$^+$, Z.\ V.\ Pchelkina$^*$, M.\ V.\ Sadovskii$^+$}

%%% author(s) - for colontitle (at the top of the page)
\rauthor{I.\ A.\ Nekrasov, Z.\ V.\ Pchelkina, M.\ V.\ Sadovskii}

%%% author(s) - for table of contents
\sodauthor{Nekrasov, Pchelkina, Sadovskii }

%%% author(s) - for table of contents
\sodauthor{Nekrasov, Pchelkina, Sadovskii }

%%% author's address(es)
\address{$^+$Institute for Electrophysics, Russian Academy of Sciences, 
Ural Division, 620016 Ekaterinburg, Russia}

\address{$^*$Institute for Metal Physics, Russian Academy of
Sciences, Ural Division, 620041 Ekaterinburg GSP-170, Russia}

%%% dates of submition & resubmition (if submitted once, second argument is *)
\dates{Today}{*}

%%% abstract
\abstract{
We present results of {\it ab initio} LDA calculations of electronic 
structure of ``next generation'' layered ironpnictide High-T$_c$ superconductor LiFeAs (T$_c$=18K).
Obtained electronic structure of LiFeAs is very similar to recently studied
ReOFeAs (Re=La,Ce,Pr,Nd,Sm) and AFe$_2$As$_2$ (A=Ba,Sr) compounds.
Namely close to the Fermi level its electronic properties are also determined mainly by Fe 
3$d$-orbitals of FeAs$_4$ two-dimensional layers. Band dispersions of LiFeAs are very similar to the
LaOFeAs and BaFe$_2$As$_2$ systems as well as the shape of the Fe-3$d$ density of states
and Fermi surface. 
}


\maketitle

Recently several series of new layered ironpnictide
superconductors with T$_c$ about 40--55 K has attracted a lot of scientific interest.
At present there are two types of such systems
(i) Re111 (Re=La,Ce,Pr,Nd,Sm) with parent compound LaO$_{1-x}$F$_x$FeAs
\cite{kamihara_08,chen,zhu,mand,chen_3790,chen_3603,ren_4234,ren_4283} and
(ii) A122 (A=Ba, Sr) with parent system BaFe$_2$As$_2$ \cite{rott,ChenLi,Chu}. 
The A122 systems are found to form large enough single crystals \cite{Bud}.
A new type of FeAs based superconductor, LiFeAs, was just obtained experimentally \cite{wang_4688}.
With Li ion deficiency for the composition Li$_{0.6}$FeAs T$_c$=18 K was observed.

Electronic structure of La111 series obtained by means of LDA were reported by several groups 
\cite{singh,dolg,mazin} and qualitatively agreed with the first one calculated for LaOFeP \cite{lebegue}.
Also it was shown for Re111 that band structure is rather irrelevant to the Re type \cite{Nekr}.
For the Ba122 first LDA density of states (DOS) were published in Refs. \cite{Shein, Krell}.
Detailed comparison of prototype systems La111 and Ba122 was performed
in Ref.~\cite{Nekr2}.

In this short note we present LDA band structure analysis for 
newly discovered ironpnictide system LiFeAs.
The obtained LDA band dispersions, DOS and Fermi surface for LiFeAs in comparison with 
the one previously reported for La111 and Ba122 (Refs.~\cite{Nekr,Nekr2}) are discussed.

\section{Crystal structure}\label{str}
Recently the crystal structure of LiFeAs was refined~\cite{cryst}.
LiFeAs crystallizes in tetragonal structure with the 
space group $P$4/$nmm$ and lattice parameters $a=3.7914(7)$~\AA, $c=6.364(2)$~\AA.
The experimentally obtained crystallographic positions are the following Fe(2b)
(0.75, 0.25, 0.5), Li(2c)  (0.25, 0.25, z$_{Li}$),
As(2c) (0.25, 0.25, z$_{As}$), z$_{As}$=0.26351,
z$_{Li}$=0.845915~\cite{cryst}.
The crystal structure of LiFeAs displayed in Fig. 1
has pronounced layered structure
presuming quasi two-dimensional electronic properties.
To some extent it resembles the crystal structure of 
La111~\cite{kamihara_08} and Ba122~\cite{rotter_4021}.
 
Most relevant interatomic distances Fe-Fe and Fe-As are 2.68 and 2.42 \AA~correspondingly.
At the moment it is spatially most compact crystal structure (see for comparison Ref.~\cite{Nekr2}).
As-Fe-As angles in the case of LiFeAs have values $\sim 103.1^\circ$ and $\sim 112.7^\circ$.
Thus one should expect some fine distinctions for LiFeAs with respect to La111 and Ba122~\cite{Nekr2}.

\begin{figure}[!h]
\includegraphics[clip=true,width=0.4\textwidth]{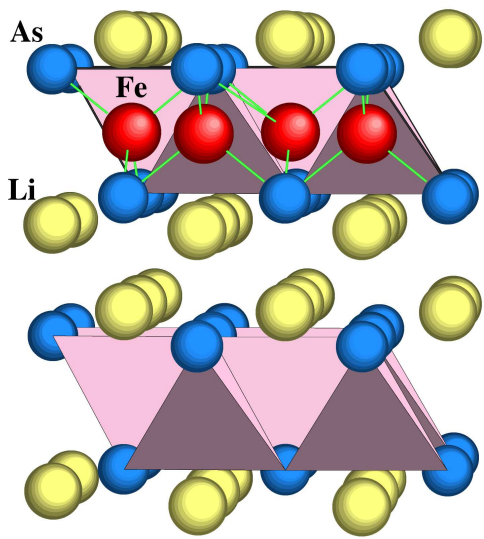}
\label{lifeas_struc}
\caption{Fig. 1. Crystal structure of LiFeAs. 
Fe ions (red) within the As (blue) tetrahedra form two-dimensional layers separated by Li ion (yellow) stratum.}
\end{figure}

\section{Electronic structure}\label{lda}
 
The electronic structure of LiFeAs compound was calculated 
within the local density approximation (LDA) by using linearized muffin-tin orbitals basis 
(LMTO)~\cite{LMTO}.
In upper part of Fig.~2 LDA calculated total, Fe-3$d$ and As-4$p$ DOS (left panel) matched with the band dispersions
(right panel) are presented. Analyzing Fig.~2 one can see that around the Fermi level from -2.5 eV to +2.5 eV
there are Fe-3$d$ states while As-4$p$ states are lower in energy from -2.5 eV down to -6.0 eV.
In the ($k_x,k_y$) plane band dispersions around the Fermi level have similar
shape as reported before for La111 and Ba112~\cite{Nekr2} (see Fig.~2, lower panel).

To compare directly LiFeAs with La111 and Ba122 materials in Fig.~3 we show
Fe-3$d$ DOS (upper panel) together with the total DOS in the vicinity of the Fermi level
(lower panel) for all of them.
A bit larger bandwidth of Fe-3$d$ states in LiFeAs is connected with more tight As tetrahedron
coordination. For the same reason As-4$p$ states are lower in energy for LiFeAs as to La111 and Ba112.
The value of total DOS on the Fermi level is 3.86 state/eV/cell (see Fig.~3 lower panel)
which is slightly less than those for La111 (4.01) and Ba122 (4.22)~\cite{Nekr2}.
In principle, this fact alone may lead to lower values
of superconducting $T_c$ in this compound, as compared with other
ironpnictide superconductors. The orbitally resolved Fe-3$d$ DOS for LiFeAs is shown in Fig.~4.
One can see three Fe-3$d$ orbitals of $t_{2g}$ symmetry -- 
$xz$, $yz$, $xy$ mainly contributing to the bands crossing the Fermi level.

In Fig. 5 we show LDA calculated Fermi surface of LiFeAs.
If compared with Fermi surfaces for Ba122 and La111 
systems again one can see two hole cylinders around $\Gamma$ point and two
electron elliptical cylinders in the corners of the Brillouin zone.
This fact can be viewed also in the band dispersions (see Fig.~2, lower panel).
However in contrast to Ba112 and La111 electron cylinders for LiFeAs are
stronger separated around A-point and have more dispersion along $k_z$ axis.

\begin{figure}
\includegraphics[clip=true,width=0.6\columnwidth,angle=270]{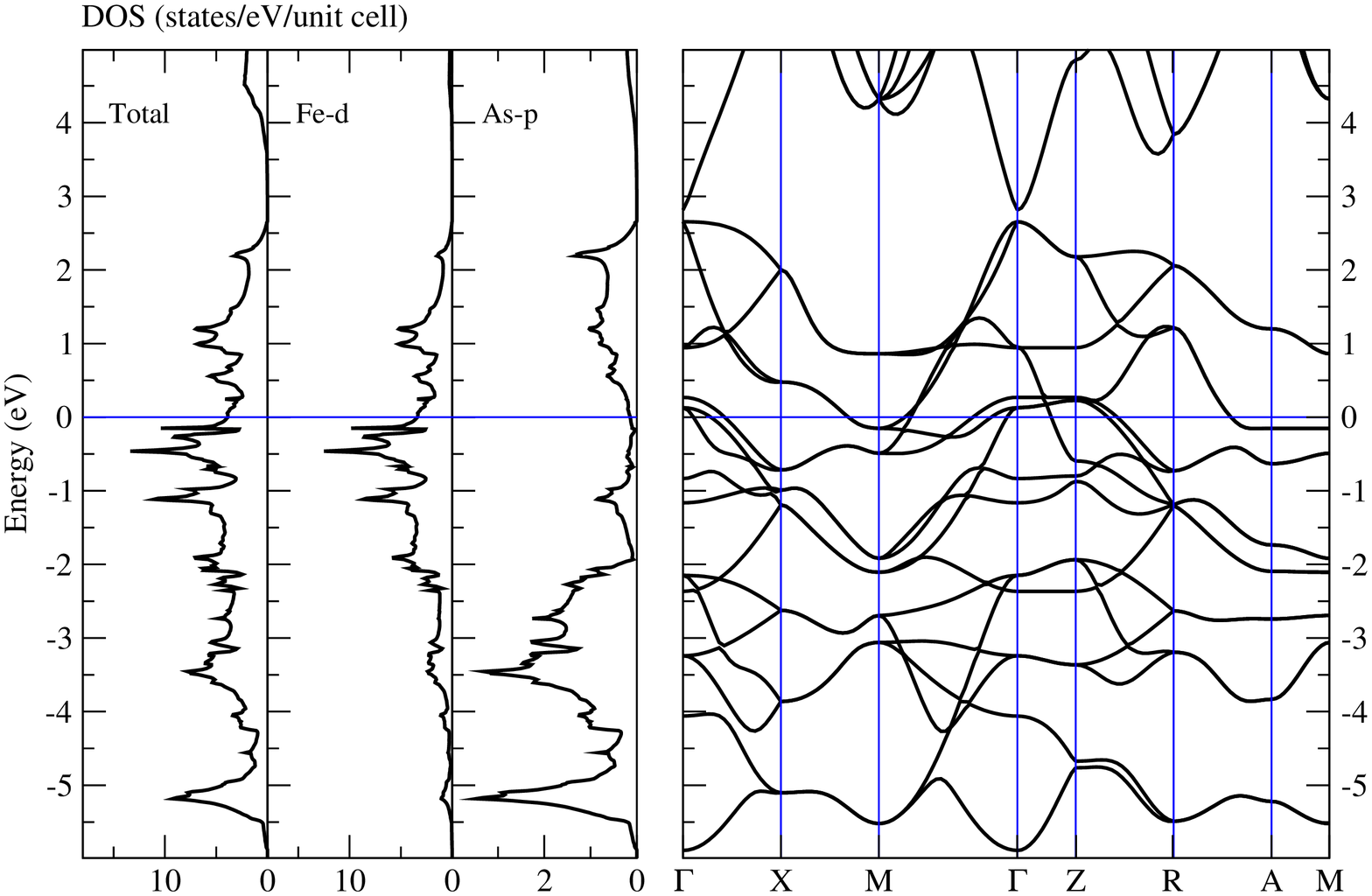}
\includegraphics[clip=true,width=0.9\columnwidth]{bands_Ef_Li.eps}
\includegraphics[clip=true,width=0.9\columnwidth]{bands_Ef_La.eps}
\includegraphics[clip=true,width=0.9\columnwidth]{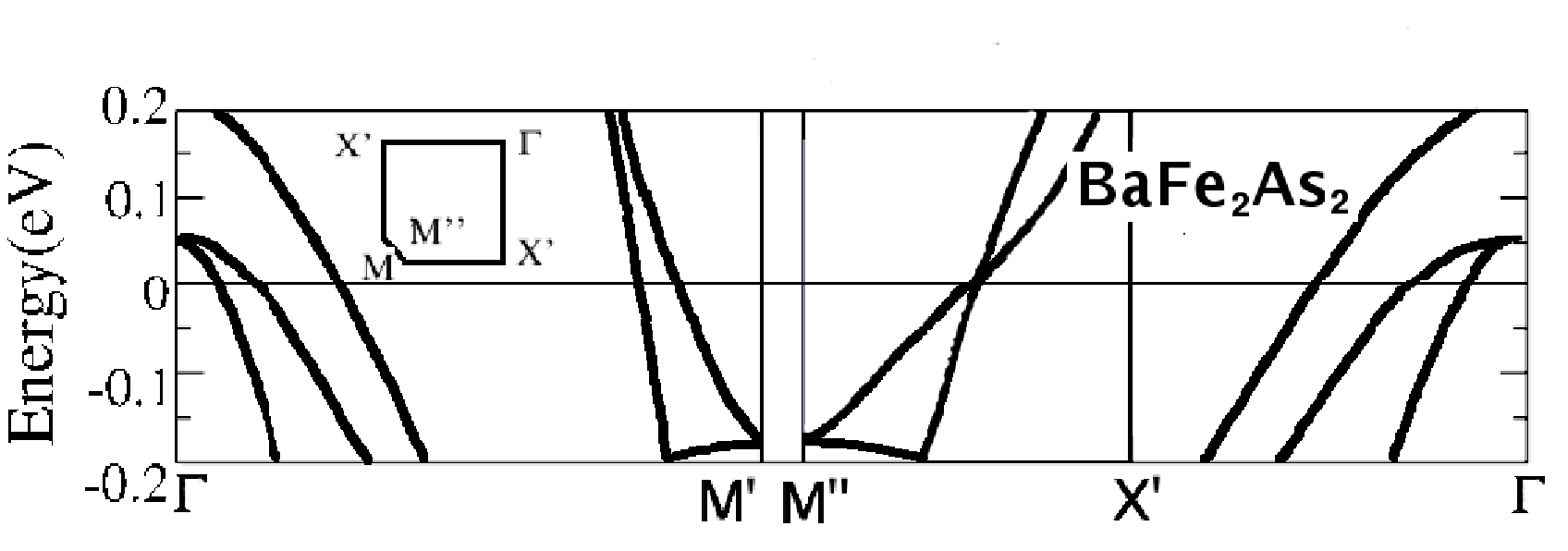}
\caption{\label{bands} 
Fig. 2. Upper panel -- LDA densities of states (left) vs. band dispersions (right) for LiFeAs.
Lower panels -- magnified band dispersions around the Fermi level for ($k_x,k_y$) plane
for LiFeAs, La111 and Ba112. The Fermi level corresponds to zero. }
\end{figure}

\begin{figure}
\includegraphics[clip=true,width=0.8\columnwidth,angle=270]{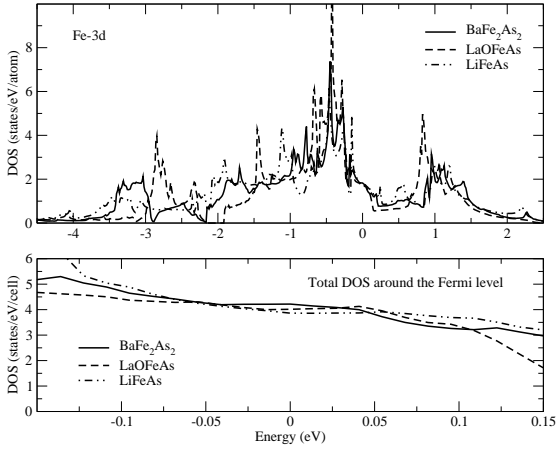}
\caption{\label{la_ba_pdos}Fig. 3. Comparison of LDA Fe-3$d$ DOS (upper panel)
for LiFeAs (dash-dotted line),
La111 (dotted line) and Ba122 (solid line).
Lower panel -- total DOS of LiFeAs, La111 and Ba122 in the vicinity of the Fermi level.
The Fermi level corresponds to zero.}
\end{figure}

\begin{figure}
\includegraphics[clip=true,width=0.9\columnwidth,angle=270]{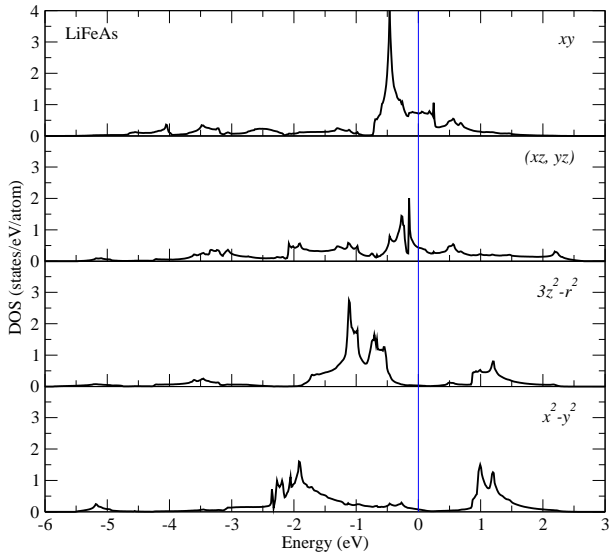}
\caption{\label{la_ba_dos}Fig. 4. Orbitally resolved LDA calculated Fe-3$d$ DOS for LiFeAs.
The Fermi level corresponds to zero.}
\end{figure}

\begin{figure}
\includegraphics[clip=true,width=0.9\columnwidth]{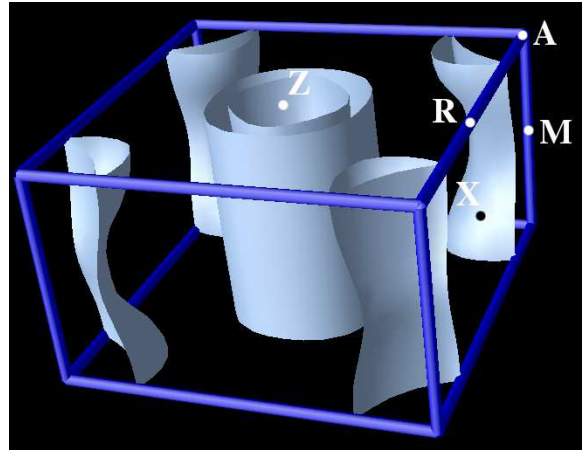}
\caption{\label{la_ba_dos}Fig. 5. LDA Fermi surface of LiFeAs
shown in the first Brillouin zone centered at $\Gamma$ point.}
\end{figure}

\section{Conclusion}

To summarize, we have presented the LDA calculation results of ``next generation''
prototype ironpnictide high-temperature superconductor LiFeAs.
A comparison of our LDA data with previously reported Re111 and A122 series 
shows resembling electronic properties. 
For all of them including LiFeAs two-dimensional FeAs$_4$ layer provides 
electronic states in the vicinity of the Fermi level which are important to 
superconductivity investigation in those compounds.

\section{Acknowledgements}

This work is supported by RFBR grants 08-02-00021, 08-02-00712, RAS programs 
``Quantum macrophysics'' and ``Strongly correlated electrons in 
semiconductors, metals, superconductors and magnetic materials'',
Grants of President of Russia MK-2242.2007.2(IN), MK-3227.2008.2(ZP)
and scientific school grant SS-1929.2008.2,  interdisciplinary 
UB-SB RAS project, Dynasty Foundation (ZP) and Russian Science Support 
Foundation(IN). The authors are grateful to I. Mazin for useful discussions.
IN thanks MPIFKS Dresden for hospitality and UB RAS for travel grant.

\end{document}